\documentclass[11pt,a4paper]{article}
\usepackage{jheppub}
\usepackage{graphicx}
\usepackage{amssymb}
\usepackage{amsmath}
\usepackage{topcapt}
\usepackage{epstopdf}
\title{Multi-matrix models at general coupling}

\author[a]{Veselin G. Filev}
\author{and}
\author[b]{Denjoe O'Connor}

\affiliation[a,b]{School of Theoretical Physics, Dublin Institute for Advanced Studies\\
10 Burlington Road, Dublin 4, Ireland.}
\emailAdd{vfilev@stp.dias.ie}
\emailAdd{denjoe@stp.dias.ie}

\abstract{The eigenvalue distribution of Hoppe's two matrix model is
  investigated in detail as a function of the model's coupling. For
  small couplings it is a perturbed Wigner semicircle, while for large
  couplings it is a parabolic distribution which crosses over to a
  Wigner semicircle for eigenvalues within approximatley an inverse
  coupling from the boundary of the distribution. The model is
  approximately commuting at large couplings and we find the joint
  eigenvalue distribution of the two matrices.  We also study a
  related three matrix model finding the corresponding three
  dimensional eigenvalue distribution there also. The techniques
  developed here are more widely applicable to other multi-matrix
  models.}


\subheader{}

\begin{document}

\begin{flushright}
DIAS-STP-13-06
\end{flushright}
\maketitle

\section{Introduction}
Multi-matrix models play an important r\^ole in several branches of
modern physics especially in matrix string theory
\cite{Dijkgraaf:1997vv}, the IKKT model \cite{Ishibashi:1996xs} (and
its lower dimensional variants \cite{Connes:1997cr}) and the BFFS and
BMN models \cite{Banks:1996vh,Berenstein:2003gb}. They also describe
the low energy dynamics of $D$-branes \cite{Kazakov:1998ji} and
provide simple models of emergent geometry
\cite{DelgadilloBlando:2007vx,DelgadilloBlando:2012xg} and emergent gravity
\cite{Steinacker:2012ra,Blaschke:2010ye}.

There are very few exactly solvable interacting multi-matrix models aside 
from Hoppe's two matrix model \cite{Hoppe:PhDThesis1982} which we analyse
in this paper. This model plays a r\^ole similar to that of the two
dimensional Ising model in critical phenomena, being exactly solvable
for many quantities while having a rich behaviour that is not easily
amenable to exact analysis. It also gives the characteristic behaviour
one can expect in a wider class of multi-matrix models.  The model was
introduced as a guide to the physics of quantised membranes by Hoppe \cite{Hoppe:PhDThesis1982} and it has subsequently arisen in the low energy 
dynamics of $D$-branes \cite{Kazakov:1998ji} and in discussions of 
emergent geometry \cite{Berenstein:2008eg}.

In \cite{O'Connor:2012vr} we studied the large coupling behaviour of
Hoppe's model. We established that the matrices are approximately
commuting at large coupling and the eigenvalue distribution of one of
the matrices obeys a parabolic distribution. In an appendix we
outlined the leading large coupling corrections to this parabolic
distribution.  We also established that the rotationally invariant
3-matrix model which reduces to Hoppe's model when one of the matrices
is integrated out, has, at large coupling, a uniform joint eigenvalue
distribution within a ball of radius
$R\simeq\left(\frac{3\pi}{2g}\right)^{\frac{1}{3}}$ where $g$ is the
coupling of the model.

In this paper we pursue a more thorough investigation of Hoppe's model
and make some further observations on its 3-matrix relative.

The principal results of this paper are:
\begin{itemize}
\item The eigenvalue distribution as a function of coupling, $g$, with
  perturbative expressions for large and small couplings.
\item The eigenvalue density of one matrix can be ``lifted'' to give
  rotationally invariant two and three dimensional distributions,
  which for large $g$ become the eigenvalue distributions for the two
  matrix model and its three matrix relative, respectively.
\item The unique rotationally invariant ``lift'' to two dimensions is,
  for large coupling, a hemispherical distribution with a finite
  eigenvalue density at the boundary.
\item The unique rotationally invariant ``lift'' to three dimensions
  is the uniform distribution \cite{O'Connor:2012vr}, however for any
  finite coupling, the ``lifted'' distribution grows in the shell
  $1/g\sim r < R$ and diverges at the boundary. When reduced to the
  one dimensional distribution this corresponds to a crossover to the
  Wigner distribution as the boundary is approached.
\end{itemize}

The structure of the paper is as follows:

In section 2 we introduce Hoppe's two matrix model, expand around a
background of diagonal matrices, gauge fix so that one linear
combination of the matrices is diagonal.  We then integrate out the
all remaining modes to obtain an effective action for these
longitudinal modes (eigenvalues of the diagonalised matrix). We then
average over the arbitrary unit vector selecting the diagonalised
matrix to get an integral equation for a rotationally invariant two
dimensional distribution. We call this distribution the two
dimensional ``lift'' of the eigenvalue distribution. The remainder of
the section deals with analysing this ``lifted'' integral equation for
weak and strong coupling.  We establish that at weak coupling the
solution is a uniform distribution while at strong coupling it gives a
hemispherical distribution. The hemispherical lifted distribution in
turn implies a parabolic eigenvalue distribution for the eigenvalue
distribution of a single matrix.

Section 3 considers the one dimensional eigenvalue distribution and
begins by developing perturbation theory around weak coupling. At zero
coupling the eigenvalue distribution is the Wigner semicircle and we
show that up to order $(R g)^6$ (where $R$ is the extent of the
distribution and $g$ the coupling) the distribution is a Wigner
semicircle modified by polynomials in $\eta=x/R$ (see equation
(\ref{rhoxapr})). Section 3.2 then develops perturbation theory for
large coupling where the leading form of the distribution at large $g$
is a parabola. An analytic form for the leading correction to the
parabolic distribution is obtained. It is then shown that this
reproduces the exact asymptotic growth of the observable
$\nu=g^2<\frac{\rm Tr}{N}(X^2)>$ ,for large $g$, as obtained from and
exact expression found in \cite{Kazakov:1998ji}.

In section 3.3 we develop a numerical technique based on the
Multhopp--Kalandiya method \cite{Polyanin} to find the eigenvalue
distribution for arbitrary couplings and verify our analytic
approximations for weak and strong couplings giving the regime of
validity of these.

Section 4 demonstrates that given a $d-1$ dimensional distribution
one can determine the rotational distribution ``lifted'' to $d$
dimensions. It establishes that the uniform distribution is the
``lift'' of the Wigner semi-circle and that lifting the parabolic
distribution with its leading correction, equation (\ref{rhoxapr}),
leads to a truncated hemispherical distribution (see figure \ref{fig:5}).

Section 5 discusses a 3 matrix variant of Hoppe's two matrix model
which at large coupling was shown, \cite{O'Connor:2012vr}, to have a
eigenvalues uniformly distribution within a solid ball of radius
$(\frac{3\pi}{2g})^{1/3}$. The integral equation for the rotationally 
invariant 3 dimensional ``lifted''  distribution is established. 
It is shown that quite generally the rotationally invariant  $d-2$ 
dimensional distribution lifts to a rotationally invariant $d$ dimensional distribution given by $\rho_d(x)=-\frac{\rho'_{d-2}(x)}{2\pi x}$. The Wigner semicircle and perturbations of it at small couplings 
lift to distributions that are divergent at the boundary, while the parabolic 
distribution lifts to the uniform distribution within a ball. The leading 
corrections to the parabola lifts to a distribution that is again divergent 
at the boundary and we establish by numerical integration of the integral equation that the characteristic behaviour for all finite couplings is the square root divergence characteristic of the ``lifted'' Wigner semicircle.

The paper has two technical appendices, the first, Appendix A, deals with the solution to the 2-dimensional ``lifted'' integral equation at large coupling while the second, Appendix B, deals with the large $g$ asymptotics of the exact radial extent of the eigenvalue distribution $R(g)$ and the large $g$ asymptotics of the observable $\nu$.

\section{The two-matrix model.}
The principal model that we focus on in these notes is the two
dimensional mass regulated model first considered by Hoppe
\cite{Hoppe:PhDThesis1982}:
\begin{equation}
{\cal Z}=\int{\cal D}X{\cal D}Ye^{-N{\rm{tr}}({X^1}^2+{X^2}^2-g^2{[X^1,X^2]}^2)}\ .\label{partfunct2}
\end{equation}
Our main interest is the properties of this model at strong coupling, when it is in a nearly commuting\footnote{For large $g$ we have
$<\frac{Tr}{N}(i[X,Y])^2>\simeq\frac{1}{2g^2}-\frac{1}{5g^2}(\frac{3\pi}{2g})^{2/3}$,
and so $X$ and $Y$ commute for $g\rightarrow\infty$.} phase \cite{Berenstein:2008eg}. The main strategy to solve the model is to reduce it to a one dimensional model, from which many properties can be extracted exactly. 

However we will first use the approach of ref. \cite{O'Connor:2012vr} and study the two matrix model directly by obtaining a two-dimensional distribution, which in the commuting phase coincides with the joint eigenvalue distribution of the matrices. To this end we split the matrices as:
\begin{equation}
X^1_{ij}=x^1_i\delta_{ij}+a^1_{ij};~~~X^2_{ij}=x^2_i\delta_{ij}+a^2_{ij};~~~\vec x_i=(x^1_i,x^2_i);~~~\vec a_{ij}=(a^1_{ij},a^2_{ij})\ .
\end{equation}
Consider a constant unit vector $\vec n=(n^1,n^2)$ and define:
\begin{equation}
\vec{x}^{||}=\vec n(\vec n.\vec x);~~~\vec{x}^{\perp}=(\hat 1-\vec n\vec n).\vec x;~~~\vec a^{||}=\vec n(\vec n.\vec a);~~~\vec a^{\perp}=(\hat 1-\vec n\vec n).\vec a\ .
\end{equation}
Now we can use the $SU(N)$ symmetry of the matrix model to fix the gauge:
\begin{equation}
\vec n.\vec a_{ij}=0\ .\label{gauge}
\end{equation}
After integrating out the perpendicular elements of the matrices $\vec{x}^{\perp}$ and $\vec a^{\perp}$ the resulting effective action for $(\vec n.\vec x)$ is \cite{O'Connor:2012vr}:
\begin{equation}
S_{\rm{eff}}[\vec x]=\frac{1}{N}\sum_{i=1}^N({\vec n.\vec x_i})^2-\frac{1}{2N^2}\sum_{i,j=1}^N\log\left[\frac{(\vec n.(\vec x_i-\vec x_j))^2}{1+g^2(\vec n.(\vec x_i-\vec x_j))^2}\right]\ .\label{eff-action}
\end{equation}
Next, we consider, for large $N$, the continuous limit of equation (\ref{eff-action}) and define a rotationally invariant two dimensional 
distribution $\rho(\vec x)$ so that (\ref{eff-action}) becomes
\begin{eqnarray}\label{rotinvact}
S_{\rm{eff}}[\rho(\vec x)]&=&\int d^2x\rho(\vec x)(\vec n.\vec x)^2-\frac{1}{2}\int\int d^2x d^2x'\rho(\vec x)\rho(\vec x')\log\left[\frac{(\vec n.(\vec x-\vec x'))^2}{1+g^2(\vec n.(\vec x-\vec x'))^2}\right]+\quad\quad\quad \\
&+&\mu\left(\int d^2x\rho(\vec x)-1\right)\nonumber\ .
\end{eqnarray}
Note that in general the matrices $X^{\mu}$ do not commute and
$\rho(\vec x)$ is a rotationally invariant ``lifted" version of the
one dimensional distribution of $\vec n.\vec x$. However $\vec n .\vec
x$ is the eigenvalue of the matrix $\vec n. \vec X$ and when the
matrices commute diagonalising $\vec n. \vec X$ would diagonalise both
$X^1$ and $X^2$. Therefore for $g^2\to \infty$ when the model is in a
commuting phase $\rho(\vec x)$ approaches the joint eigenvalue
distribution of $X^{\mu}$. At weak coupling the model is non-commuting
and the ``lifted" two dimensional distribution is a rotationally
invariant lift of the eigenvalue distribution of one of the matrices,
and is not itself an eigenvalue distribution.

Varying with respect $\rho$ in equation (\ref{rotinvact}) we obtain:
\begin{equation}
\mu+(\vec n.\vec x)^2=\int d^2x'\rho(\vec x')\log\left[\frac{(\vec n.(\vec x-\vec x'))^2}{1+g^2(\vec n.(\vec x-\vec x'))^2}\right]\ .\label{EQN}
\end{equation}
Note that equation (\ref{EQN}) should be valid for any choice of $\vec n$ thus in order to obtain a rotationally invariant integral equation we average over $\vec n$ with weight one.\footnote{For $\vec n =(\cos\phi,\sin\phi)$  we integrate both sides of the equation by $\frac{1}{2\pi}\int\limits_0^{2\pi}d\phi$\ .} The result is:
\begin{equation}
\mu+\frac{\vec{x}^2}{2}=2\int d^2x'\rho(\vec{x'})\ln(\frac{\vert\vec{x}-\vec{x'}\vert}{1+\sqrt{1+g^2\vert\vec{x}-\vec{x'}\vert^2}}) .\label{ROTINVEQN}
\end{equation}
\subsection{2D distribution at weak coupling}

To obtain an integral equation suitable for
perturbative calculation at small $g$ we apply 
$\vec\nabla_x^2$ on both sides of equation (\ref{rotinvact}). We
obtain:
\begin{eqnarray}
2&=&\int d^2 x'\,\rho(\vec x')\,\left[ \vec\nabla_x \left(\frac{2}{\sqrt{1+g^2(\vec x-\vec x')^2}}\right).\frac{\vec x-\vec x'}{|\vec x-\vec x'|^2} +\frac{4\pi}{1+\sqrt{1+g^2\vert\vec{x}-\vec{x'}\vert^2}}\delta(\vec x-\vec x')\right]\nonumber\\
&=&4\pi\rho(\vec x)-2g^2\int\,d^2x'\frac{\rho(\vec x')}{(1+g^2(\vec x-\vec x')^2)^{3/2}}\ ,\label{inteq2kind}
\end{eqnarray}
which is an integral equation of the second kind. To avoid complications with the boundary of the integral (since the radius of the distribution runs with $g$) it is convenient to introduce a new variable $\vec\eta=\vec x/R$. Equation (\ref{inteq2kind}) can then be written as:
\begin{equation}
\rho(\vec\eta)=\frac{1}{2\pi}+\frac{(Rg)^2}{2\pi}\int\limits_{|\eta'|\leq 1} d^2 \eta'\frac{\rho(\vec \eta')}{\left(1+(Rg)^2(\vec \eta-\vec \eta')^2\right)^{3/2}}\ .\label{equationeta||}
\end{equation}
The kernel in equation (\ref{equationeta||}) is bounded by:
\begin{equation}
K_2(Rg,|\vec \eta-\vec \eta'|)\equiv\frac{1}{2\pi}\int\limits_0^{2\pi}d\phi \frac{\eta'}{\left(1+(Rg)^2(\vec \eta-\vec \eta')^2\right)^{3/2}}\leq1 \ ,
\end{equation}
therefore for sufficiently small $(Rg)^2$ we can solve equation (\ref{equationeta||}) iteratively:
\begin{equation}
 \rho(\vec\eta)=\frac{1}{2\pi}\left[1+(Rg)^2\int d^2\eta' K_2+(Rg)^4\int d^2\eta' K_2\int d^2\eta'' K_2 +\dots\right]\ .\label{iterations}
\end{equation}
Using equation (\ref{iterations}) one can solve for $\rho(\eta)$ perturbatively to arbitrary order in $(Rg)$. Here we present the solution to sixth order:
\begin{equation}
\rho(\eta)=\frac{1}{2\pi}+\frac{{(Rg)}^2}{4 \pi }-\frac{\left(6 \eta ^2+1\right) {(Rg)}^4}{16 \pi }+\frac{\left(15 \eta ^4+24 \eta ^2-2\right) {(Rg)}^6}{32 \pi }+O\left((Rg)^8\right)\ . \label{2Dsmallg}
\end{equation} 
Note that $R$ in equation (\ref{2Dsmallg}) is $g$ dependent and
expansion in $g$ would look different. We can determine the $g$
dependence of $R$ using the definition of $\eta$ and the normalization
of $\rho(x)=\rho(\eta)$. Indeed:
\begin{equation}
\int d^2x\rho(x)=R^2\,\int d^2\eta \rho(\eta)=1 \label{Norm2D}
\end{equation}
We can now substitute equation (\ref{2Dsmallg}) in equation (\ref{Norm2D}), expand $R$ in terms of $g^2$ and determine the coefficients in the expansion by solving equation (\ref{Norm2D}) order by order. This procedure can be performed to arbitrary order. Here we present the result to sixth order in $g$:
\begin{equation}
R=\sqrt{2} - \frac{1}{\sqrt{2}}\,g^2 + \frac{15}{4\sqrt{2}}\,g^4 - \frac{165}{8\sqrt{2}}\,g^6+O\left (g^8\right)\ . \label{Rofsmallg}
\end{equation}

\subsection{2D distribution at strong coupling}

Next we focus on the large $g$ limit of the model. It is convenient to modify equation (\ref{ROTINVEQN}) to:
\begin{equation}
\mu'+\frac{\vec{x}^2}{2}=2\int d^2x'\rho(\vec{x'})\ln(\frac{g\vert\vec{x}-\vec{x'}\vert}{1+\sqrt{1+g^2\vert\vec{x}-\vec{x'}\vert^2}}) .\label{ROTINVEQNMOD}\ ,
\end{equation}
where the factor of $g$ on the right-hand side of the equation is compensated by a redefinition of the constant $\mu\to\mu'$. It is a straightforward exercise to obtain the asymptotic form of equation (\ref{ROTINVEQNMOD}) in the $g\to\infty$ limit:
\begin{equation}
\mu'+\frac{\vec{x}^2}{2}=-\frac{2}{g}\int d^2x'\frac{\rho(\vec{x'})}{\vert \vec x-\vec x'\vert}+O\left({1}/{g^2}\right)\ .
\end{equation}
To leading order the integral equation that we obtain is:
\begin{equation}
\mu'+\frac{\vec{x}^2}{2}=-\frac{2}{g}\int d^2x'\frac{\rho(\vec{x'})}{\vert \vec x-\vec x'\vert}=-\frac{2}{g}\int\limits_0^R dx' x'\int\limits_0^{2\pi}d\phi\frac{\rho(x')}{\sqrt{x^2+x'^2-2x x'\cos\phi}}\ ,
\end{equation}
where we have used $\rho(\vec x)=\rho(\vert\vec x\vert)=\rho(x)$. After performing the integral over $\phi$ we arrive at the integral equation:
\begin{equation}
\mu'+\frac{\vec{x}^2}{2}=-\frac{8}{g}\int\limits_0^R\,dx' \frac{x'\rho(x')}{x+x'}K\left(\frac{2\sqrt{xx'}}{x+x'}\right)\ , \label{INTEQ}
\end{equation}
where $K(z)$ is the complete elliptic integral of the first kind. The integral equation (\ref{INTEQ}) can be solved \cite{Zabreyko} (see also Appendix A) for $\rho(x)$:
\begin{equation}
\rho(x)=\frac{g}{\pi^2}\frac{\frac{1}{2}(R^2-\mu')-x^2}{\sqrt{R^2-x^2}}=\frac{g}{\pi^2}\sqrt{R^2-x^2}\ , \label{hemisphere}
\end{equation}
where we have fixed the constant $\mu'=-R^2$ by demanding that the distribution be finite at the boundary ($x=R$). Equation (\ref{hemisphere}) is the hemisphere distribution of ref. \cite{Berenstein:2008eg}. By normalizing the distribution $\rho(x)$ to one we can fix the radius of the distribution $R$:
\begin{equation}
R=\left(\frac{3\pi}{2g}\right)^{1/3}\ . \label{radius}
\end{equation}
Having obtained the hemisphere distribution for $g\to\infty$ directly in two dimensions we are interested in the behaviour of the model for finite values of the coupling constant $g$, when the model is nearly commuting. Note that strictly speaking the correction to the joint eigenvalue distribution of the model at finite coupling is not well defined since the model is not in a commuting phase. However, there is a complex observable $\Phi =X^1+i X^2$, which has complex eigenvalues which are well defined at any coupling. Furthermore in the commuting phase the real and imaginary components of the eigenvalues of $\Phi$ coincide with the components of the joint eigenvalues $\vec x$. Numerically one can simulate the model at finite $g$ and obtain the distribution of $\Phi$ keeping in mind that in the commuting phase this is the joint eigenvalue distribution. In order to compare to numerical simulations we need to understand the behaviour of the distribution for large but finite $g$. 

It turns out that it is technically easier to determine the correction to the hemisphere distribution by integrating out one of the matrices and study the corresponding one dimensional distribution. In the next section we analyze the reduced model and the one dimensional distribution at general coupling $g$. As we show by solving an integral equation of Abel's type we can lift the one dimensional distribution to a rotationally invariant two dimensional one.

\section{The one matrix model}

In this section we focus on the one-dimensional distribution of the matrix $\vec  n. \vec X$ defined in the previous section. Without loss of generality we can choose $\vec n =(1 ,0)$ and $\vec x = (x,y)$. The integral equation for the distribution $\rho_1(x)$ is given by:
\begin{equation}
\mu'+x^2=\int dx'\rho_1(x')\log\left[\frac{g^2(x-x')^2}{1+g^2(x-x')^2}\right]\ ,\label{EQN1}
\end{equation}
where we have substituted the definition of $\rho_1(x)$:
\begin{equation}
     \rho_1(x)=\int\limits_{-\sqrt{R^2-x^2}}^{\sqrt{R^2-x^2}}\rho(\sqrt{x^2+y^2})dy\ . \label{2to1}
\end{equation}
in equation (\ref{EQN}) and have redefined the constant $\mu \to \mu'$ to add the factor of $g^2$ in the argument of the logarithmic function in equation (\ref{EQN1}). It is convenient to differentiate equation (\ref{EQN1}) with respect to $x$. The resulting integral equation can be written as:
\begin{equation}
-x=\int\limits_{-R}^R dx ' \,\frac{\rho_1(x')}{x'-x}+\int\limits_{-R}^Rdx'\,\rho_1(x')\,K(g,x'-x) \ , \label{EQNKERN}
\end{equation}
where the kernel $K(g,u)$ is given by:
\begin{equation}
K(g,u)=-\frac{g^2u}{1+g^2u^2}\ . \label{KERNEL}
\end{equation}
{\subsection{1D distribution at weak coupling }
At $g=0$ we have $K(0,u)=0$ and the integral equation (\ref{EQNKERN}) has a simple Cauchy kernel:
\begin{equation}
-x=\int\limits_{-R}^R dx '\, \frac{\rho_1(x')}{x'-x}\ .\label{preWigner}
\end{equation}
Note that at vanishing coupling the model is Gaussian and hence the one dimensional distribution of $X^1$, or equivalently $\vec n.\vec X$, should be a Wigner semi-circle. The unique bounded solution of equation (\ref{preWigner}) is indeed the Wigner semicircle (\ref{Wigner}).  
\begin{equation}
     \rho_1(x)=\frac{1}{\pi}\sqrt{R^2-x^2}\ ,\label{Wigner}
\end{equation}
with radius $R=\sqrt{2}$, which agrees with the $Rg\to 0$ limit of equation (\ref{Rofsmallg}). 
The perturbative solution for small $g$ can be obtained, in terms of the coordinate $\eta=x/R$ and the coupling $Rg$, by integrating out one of the components of $\vec\eta$ in equation (\ref{2Dsmallg}) :
\begin{equation}
    \tilde \rho(\eta_1)=\int\limits_{-\sqrt{1-\eta_1^2}}^{\sqrt{1-\eta_1^2}}\rho(\sqrt{\eta_1^2+\eta_2^2})\, d\eta_2\  \ ,
\end{equation}
where $\tilde\rho(\eta)$ is related to $\rho_1(x)$ via:
\begin{equation}
\tilde\rho(\eta)=\frac{1}{R}\rho_1(R\,\eta)\ .
\end{equation}
The final expression for small $Rg$ up to sixth order is given by:
\begin{equation}
\rho_1(\eta)=\sqrt{1-\eta^2} \left[\frac{1}{\pi }+\frac{(Rg)^2}{2 \pi } -\frac{\left(4 \eta ^2+3\right) (Rg)^4}{8 \pi } +\frac{\left(8 \eta ^4+20 \eta ^2+9\right) (Rg)^6}{16 \pi }+O\left((Rg)^8\right) \right] \ . 
\label{oneDpert}
\end{equation}
As one can see, the small $Rg$ corrections deform the Wigner 
semicircle but it still has the characteristic $\sqrt{1-\eta^2}$ behaviour. 

\subsection{1D distribution at strong coupling}
For large $g$ one can use:
\begin{equation}
\frac{1}{u}+K(g,u)=-\frac{\pi}{g}\delta'(u)+O\left( 1/g^2 \right)\ 
\end{equation}
to obtain:
\begin{equation}
-x= \frac{\pi}{g}\rho_1(x)+O\left(1/g^2\right)\ ,
\end{equation}
which to leading order in $g$ is solved by the parabolic distribution \cite{Berenstein:2008eg}:
\begin{equation}
\rho_1(x)=\frac{3}{4R^3}(R^2-x^2)=\frac{g}{2\pi}(R^2-x^2)\ . \label{parabola}
\end{equation}
with radius given by equation (\ref{radius}).

In order to obtain the corrections to the parabolic distribution at finite $g$ we will derive an integral equation of the second kind which can (at least in principal) be solved iteratively. Let us begin by noting that the kernel of the integral equation (\ref{EQNKERN})  can be written as:
\begin{equation}
\frac{1}{x'-x}+K(g,x'-x)=\frac{d}{dx'}K_1(g,x'-x)\ ,
\end{equation}
where $K_1(g,x'-x)$ is the symmetric kernel:
\begin{equation}
K_1(g,x'-x)=\frac{1}{2}\log\left[\frac{g^2(x-x')^2}{1+g^2(x-x')^2}\right]\ .\label{K1}
\end{equation}
After integration by parts, and noting that $\rho(\pm R)=0$, 
equation (\ref{EQNKERN}) can be written as:
\begin{equation}
x=\int\limits_{-R}^{R}\,dx'\,K_1(g,x'-x)\,\rho_1'(x')\ . \label{eqnK1}
\end{equation}
Again, to deal with the $g$ dependence of the limits of the integral,
it is convenient to rewrite the integral equation (\ref{eqnK1}) in
terms of the variables $\eta=x/R$, $Rg$ and the distribution
$\tilde\rho(\eta)=\frac{\rho_1}{R}$ so that (\ref{eqnK1}) becomes:
\begin{equation}
\eta=\int\limits_{-1}^{1}\,d\eta'\,K_1(Rg,\eta'-\eta)\,\tilde\rho'(\eta')\ . \label{eqnK1eta}
\end{equation}
At large $Rg$ the kernel $K_1$ has the expansion:
\begin{equation}
K_1 = -\frac{\pi} {Rg}\delta(\eta'-\eta)+O\left(1/(Rg)^2\right) \label{expansionK1}\ .
\end{equation}
Next we define the kernel:
\begin{equation}
\Delta K(Rg,\eta'-\eta)=-\frac{Rg}{\pi}K_1(Rg,\eta'-\eta)-\delta(\eta'-\eta)
\end{equation}
with $\Delta\tilde\rho$ defined by:
\begin{equation}
\Delta\tilde\rho' (\eta)=\tilde\rho'(\eta)-Rg\,\tilde\rho_{(0)}(\eta)=\tilde\rho'(\eta)+\frac{Rg}{\pi}\,\eta \label{deltarho}\ ,
\end{equation}
where $\tilde\rho_{(0)}$, given by:
\begin{equation}
\tilde\rho_0(\eta)=\frac{1}{2\pi}(1-\eta^2)\label{rho0} ,
\end{equation}
is the parabolic distribution (\ref{parabola}) valid in the strict $g\to \infty$ ($Rg\to\infty$) limit. The integral equation (\ref{eqnK1eta}) can be written as:
\begin{equation}
\Delta\tilde\rho'(\eta)=-Rg\,\int\limits_{-1}^{1}d\eta' \,\Delta K(Rg,\eta'-\eta)\,\tilde\rho_{(0)}'(\eta')-\int\limits_{-1}^{1}d\eta' \,\Delta K(Rg,\eta'-\eta)\,\tilde\Delta\rho'(\eta')\ . \label{IntEqDeltaRho}
\end{equation}
Equation (\ref{IntEqDeltaRho}) is an integral equation of the second
kind for the correction $\Delta\tilde\rho$. Furthermore, from the
definition of $\Delta K$ and the expansion of $K_1$ at large $Rg$,
equation (\ref{expansionK1}), it follows that $\Delta K$ dies out at
large $Rg$ and hence the integral equation (\ref{IntEqDeltaRho}) can
be developed recursively in a convergent series:
\begin{equation}
\Delta\tilde\rho'(\eta)=-Rg\,\int \Delta K\,\tilde\rho_{(0)}' +Rg\,\int \Delta K\int \Delta K\,\tilde\rho_{(0)}' -Rg\,\int \Delta K\int \Delta K\int \Delta K\,\tilde\rho_{(0)}'+\dots\ .\label{series}
\end{equation}
If we define $\tilde\rho_{(1)}'(Rg,\eta)=-Rg\,\int \Delta K\,\tilde\rho_{(0)}'$ and:
\begin{equation}
\tilde\rho_{(n+1)}'(Rg,\eta)=-\,\int\limits_{-1}^{1} d\eta\,\Delta K(Rg,\eta'-\eta)\,\tilde\rho_{(n)}'(Rg,\eta')~~~\,n=1,2,\dots \ , \label{rhon}
\end{equation}
we arrive at the following expression for $\Delta\tilde\rho(\eta)$: 
\begin{equation}
\Delta\tilde\rho(\eta)=\sum\limits_{n=1}^{\infty}\,\int\limits_{-1}^{\eta}d\eta'\,\,\frac{\tilde\rho_{(n)}'(Rg,\eta')}{(Rg)^{n-1}}\ . \label{deltarho}
\end{equation}
At large $Rg$ we have  $1/Rg<\int\Delta K<1$. Therefore:
\begin{equation}
\frac{1}{(Rg)^{n-1}}<\frac{\tilde\rho'_{(n)}(Rg,\eta)}{(Rg)^{n-1}}<\frac{1}{(Rg)^{n-2}} \label{comprhopr}
\end{equation}
and naively one would expect that at large $Rg$ the contribution to $\Delta\tilde\rho$ in equation (\ref{deltarho}) from terms with $n>1$ would die out. However one can show that $\tilde\rho'_{(n)}(\infty,\eta)$ is not integrable near the boundary ($\eta=\pm 1$) and by regulating it with a cutoff of the order $\sim 1/(Rg)$ one can estimate that for large $Rg$:  $\int\tilde\rho'_{(n)}(Rg,\eta)\sim (Rg)^{n-1}$ for $n>1$ and $\int\tilde\rho'_{(1)}(Rg,\eta)\sim \log(Rg)$ for $n=1$. Therefore for $n>1$ we have that $\tilde\rho'_{(n)}(Rg,\eta)/(Rg)^{n-1}\sim [\delta (1-\eta)-\delta(1+\eta)]$ for sufficiently large $Rg$ and all terms with $n>1$ give a constant contribution $\kappa\sim1$ to $\Delta\tilde\rho$ in equation (\ref{deltarho}) as long as $\eta\in(-1,1)$. At the boundary the contribution from all terms vanishes and we have $\Delta\tilde\rho(\pm1)=0$. Therefore to leading order we have the following expression for $\Delta\tilde\rho$:
\begin{equation}
\Delta\tilde\rho(\eta) = \begin{cases} \int\limits_{-1}^{\eta}\,d\eta'\tilde\rho_{(1)}'(Rg,\eta')+\kappa+O((\log(Rg)/Rg)\, & \mbox{if } -1\leq\eta\leq1\\
0  & \mbox{if } ~~~~\eta=\pm1 \end{cases}\label{approxDeltarho}
\end{equation}
Using the definitions from equations (\ref{rho0}) and (\ref{rhon}) we can obtain the following expression for $\tilde\rho'_{(1)}$:
\begin{eqnarray}
\tilde\rho_{(1)}'(Rg,\eta)&=&\frac{Rg\,\eta}{\pi^2}\left[\tan^{-1}[Rg(1+\eta)]+\tan^{-1}[Rg(1-\eta)]-\pi \right]+\frac{1}{4\pi^2}\log\left[\frac{1+(Rg)^2(1-\eta)^2}{1+(Rg)^2(1+\eta)^2} \right]\nonumber\\
&+&\frac{(Rg)^2}{4\pi^2}\,(1-\eta^2)\,\log\left[\frac{(1+\eta)^2(1+(Rg)^2(1-\eta)^2)}{(1-\eta)^2(1+(Rg)^2(1+\eta)^2)}\right]\ .\label{rho1pr}
\end{eqnarray}
Note that:
\begin{equation}
\tilde\rho_{(1)}'(\infty,\eta)=-\frac{1}{\pi^2}\frac{\eta}{1-\eta^2}+\frac{1}{2\pi^2}\log\left[\frac{1-\eta}{1+\eta}\right] \ ,
\end{equation}
which is indeed not integrable near $\eta=\pm 1$. Let us introduce a cutoff $\epsilon$. The regulated expression for $\tilde\rho_{(1)}$ is:
\begin{equation}
\tilde\rho_{(1)}^{(\epsilon)}(\eta)=\frac{1}{2\pi^2}\,\eta\,\log\left[\frac{1-\eta}{1+\eta}\right]+\frac{1}{2\pi^2}\log\left[\frac{2}{\epsilon}\right]+O\left(\epsilon\,\log(\epsilon)\right)\ . \label{rho1eps}
\end{equation}
The cutoff $\epsilon$ can be expressed in terms of $Rg$. Indeed if we integrate directly the integrable expression (\ref{rho1pr}) we obtain:
\begin{equation}
\tilde\rho_{(1)}(Rg,\eta)=\frac{Rg}{2\pi}\left[T_1(Rg,\eta)+T_2(Rg,\eta)+T_3(Rg,\eta)\right] \ ,
\end{equation}
where $T_1$, $T_2$ and $T_3$ are given by:
\begin{multline}
T_1(Rg,\eta)=(1-\eta^2)\left(1-\frac{\tan^{-1}(Rg(1-\eta))+\tan^{-1}(Rg(1+\eta))}{\pi}   \right)\\ 
+\frac{\tan^{-1}(2Rg)-\tan^{-1}(Rg(1+\eta))-\tan^{-1}(Rg(1-\eta))}{3\pi\,(Rg)2}\ ,
\end{multline}
\begin{gather}
T_2(Rg,\eta)=\frac{1}{2\pi\,Rg}\left(\eta\,\log\left[\frac{1+(Rg)^2(1-\eta)^2}{1+(Rg)^2(1+\eta)^2}  \right]+\log(1+4(Rg)^2)    \right) \ ,~~~~~~~~~~~~~~~
\end{gather}
\begin{multline}
T_3(Rg,\eta)=\frac{Rg}{3\pi}\log\left[ \frac{(Rg)^4(1-\eta^2)^2}{(1+(Rg)^2(1-\eta)^2)(1+(Rg)^2(1+\eta)^2)} \right]\\
-\frac{Rg}{2\pi}\,\eta\,\left(1-\frac{\eta^2}{3}\right)\,\log\left[\frac{(1-\eta)^2(1+(Rg)^2(1+\eta)^2)}{(1+\eta)^2(1+(Rg)^2(1-\eta)^2)}   \right]\\
-\frac{2 Rg}{3\pi}\log\left[\frac{4(Rg)^2}{1+4(Rg)^2}\right]\ .
\end{multline}
For the large $Rg$ expansion of $\tilde\rho_{(1)}$ we obtain:
\begin{equation}
\tilde\rho_{(1)}(Rg,\eta)=\frac{1}{2\pi^2}\,\eta\,\log\left[\frac{1-\eta}{1+\eta}\right]+\frac{1}{2\pi^2}\,\log(2\,e^{3/2}\,Rg)+O(\log(Rg)/(Rg))\ . \label{rho1exp}
\end{equation}
Comparing equations (\ref{rho1eps}) and (\ref{rho1exp}) we conclude
that $\epsilon = e^{-3/2}/(Rg)\sim1/(Rg)$, which agrees with the
analysis performed below equation (\ref{comprhopr}). Taking into
account the constant $\kappa$ from equation (\ref{approxDeltarho}) we
arrive at the following expression for $\tilde\rho$ in the
interval $\eta\in(-1,1)$:
\begin{equation}
\tilde\rho(\eta)=\frac{Rg}{2\pi}\,\left(1-\eta^2\right)+\frac{1}{2\pi^2}\,\eta\,\log\left[\frac{1-\eta}{1+\eta}\right]+\frac{1}{2\pi^2}\,\log(2\,e^{3/2}\,Rg)+\kappa+O(\log(Rg)/(Rg))\ . \label{Deltarhokap}
\end{equation}
Note that at large $Rg$ the correction, $\Delta\tilde\rho$, vanishes
at $\eta=1-\delta_1$, where to leading order $\delta_1
=\frac{W(-3/2-2\pi^2\kappa)}{2\,\pi\,Rg}$ and $W(z)$ is the Lambert's product 
log function --- the solution to $z =W(z)e^{W(z)}$.  Therefore, at large $Rg$, to a very
good approximation $\Delta\tilde\rho(\eta)$ is given by equation
(\ref{Deltarhokap}) in the interval $\eta\in(-1+\delta_1,1-\delta_1)$
and can be taken as zero outside this interval.

The constant $\kappa$ in equation (\ref{Deltarhokap}) can be
determined numerically by computing $\tilde\rho_{(n)}$ up to
sufficiently large $n$. However one can also fix $\kappa$ indirectly
by comparing to some of the exact relations for the observables of
this model. Indeed, following the approach of
ref.~\cite{Kazakov:1998ji} one can obtain an exact relation between
the radius of the distribution and the coupling constant $g$. This
relation is rather complex and is given parametrically in terms of
elliptic integrals.  For a more detailed derivation we refer the reader
to appendix B. Here we provide only the first few terms in the large
$g$ expansion of $R$.
\begin{equation}
R=\left(\frac{3\pi}{2}\right)^{1/3}\,g^{-1/3}-\frac{2\log\,g+\log(96\pi^4)}{6\pi}g^{-1}+O(g^{-5/3})\ .\label{Rofglargeg}
\end{equation}
The radius of the distribution $R$ can also be calculated using that:
\begin{equation}
\frac{1}{R^2}=\int\limits_{-1}^{1}d\,\eta\,\tilde\rho(\eta)=\frac{Rg}{2\pi}\int\limits_{-1}^{1}\,d\eta\,(1-\eta^2)+\int\limits_{-1}^{1} d\eta\,\Delta\tilde\rho(\eta)\ .\label{eqnRrho}
\end{equation}
Substituting the expression for $\Delta\tilde\rho$ from equation (\ref{Deltarhokap}) into equation (\ref{eqnRrho}) and solving for $R$ order by order in $g$ one arrives at:
\begin{equation}
R=\left(\frac{3\pi}{2}\right)^{1/3}\,g^{-1/3}-\frac{4\log\,g+2\log(12\pi)+12\pi^2\,\kappa+3}{12\pi}g^{-1}+O(g^{-5/3})\ .\label{Rofgrho}
\end{equation}
Comparing equations (\ref{Rofglargeg}) and (\ref{Rofgrho}) we obtain:
\begin{equation}
\kappa=\frac{\log(4\pi^2)-1}{4\pi^2}\ .
\end{equation}
and we find the correction $\Delta\tilde\rho$ is given by:
\begin{equation}
\Delta\tilde\rho(\eta)=\frac{1}{2\pi^2}\,\eta\,\log\left[\frac{1-\eta}{1+\eta}\right]+\frac{\log(4\pi\,Rg)+1}{2\pi^2}+O\left(\frac{\log(Rg)}{Rg}\right)
\label{DeltaRho}
\end{equation}
and our expression for the $\tilde\rho(\eta)$ in the large $Rg$ regime is
\begin{equation}
\tilde\rho(\eta)\simeq\begin{cases}\frac{Rg}{2\pi}(1-\eta^2)+\frac{1}{2\pi^2}\,\eta\,\log\left[\frac{1-\eta}{1+\eta}\right]+\frac{\log(4\pi\,Rg)+1}{2\pi^2}
+O\left(\frac{\log(Rg)}{Rg}\right) & \mbox{if } |\eta|\leq1-\delta\\
0  & \mbox{if } |\eta|\geq1-\delta \end{cases}\ ,
\label{rhotildeLarge}
\end{equation}
where $\delta =W(1/e)/(2\pi Rg)\sim 1/Rg$ and $W(z)$ is the Lambert's product 
log function.

Our expression for the leading correction to the distribution can be tested by calculating the observable $\nu=g^2<Tr X^2>$, which was obtained in closed form in ref.~\cite{Kazakov:1998ji} (see also appendix B). Indeed from the definition of $\nu$ and $\tilde\rho(\eta)$ it follows that:
\begin{equation}
\nu=g^2\,\int\limits_{-R}^{R}dx\,x^2\rho_1(x)=R^2(Rg)^2\int\limits_{-1}^{1}d\eta\eta^2\tilde\rho(\eta)=\frac{(12\pi)^{2/3}}{20}g^{4/3}-\frac{3}{(12\pi)^{2/3}}g^{2/3}+O(g^0)\ ,\label{nuofg2}
\end{equation}
which is in perfect agreement with the result of ref.~\cite{Kazakov:1998ji}. In deriving (\ref{nuofg2}) we have used equations (\ref{Rofglargeg}) and (\ref{rhotildeLarge}). 

Finally let is obtain the leading order behavior of $\rho_1(x)$ at large $g$. Using $\rho_1(x)=R\tilde\rho(x/R)$ and equations (\ref{Rofglargeg}) and (\ref{rhotildeLarge}) we obtain:
\begin{equation}
\rho_1(x)=\frac{g}{2\pi}\left[\left(\frac{3\pi}{2g}\right)^{2/3}-x^2\right]+\frac{x}{2\pi^2}\log\left[\frac{\left(\frac{3\pi}{2g}\right)^{1/3}-x}{\left(\frac{3\pi}{2g}\right)^{1/3}+x} \right]+\frac{1}{2\pi^2}\left(\frac{3\pi}{2g}\right)^{1/3}+O\left(\frac{\log g}{g}\right) 
\label{rhoxapr}
\end{equation}

In the next subsection we develop a numerical routine to obtain the one dimensional distribution for arbitrary values of the coupling $(Rg)$.

\subsection{Interpolating solution for general coupling.}

In this subsection we construct an interpolating solution to the integral equation:
\begin{equation}
-\eta=\int\limits_{-1}^1 d\eta'\, \frac{\tilde\rho(\eta')}{\eta'-\eta}+\int\limits_{-1}^1d\eta'\,\tilde\rho(\eta')\,K(Rg,\eta'-\eta) \ , \label{EQNETA}
\end{equation}
where $K$ is defined in equation (\ref{KERNEL}). Equation (\ref{EQNETA}) is a singular integral equation of the first kind with a Cauchy kernel. An interpolating solution of this equation can be found using the Multhopp--Kalandiya method (see ref.\cite{Polyanin} Chapter 14.5). When the source is odd ($-\eta$ in our case)  the approximate solution bounded at the boundary of the distribution ($\eta=\pm 1$) is given by:
\begin{equation}
\tilde\rho_a(\cos\theta)=\frac{4}{\pi(2n+1)}\sum\limits_{l=1}^{n}\tilde\rho(\cos\theta_l)\frac{\cos\left(\frac{l\pi}{2n+1}\right)\,\sin\left(\frac{n\,l\,\pi}{2n+1}\right)\,\sin((2n+1)\theta)}{\cos(2\theta)-\cos\left(\frac{2l\pi}{2n+1}\right)},
\end{equation}
where $\cos\theta=\eta$, $2n$ is the number of nodes into which the interval
$[-1,1]$ is divided and $\theta_l$ are related to the roots of the
Chebyshev polynomial of the second kind:
\begin{equation}
\eta_l=\cos\theta_l,~~\theta_l=\frac{l\pi}{2n+1}\ ,~~~l=1,\dots , 2n \ .
\end{equation}
The value of $\tilde\rho$ at the nodes $\eta_l$ is determined by the system of linear algebraic equations:
\begin{equation}
\sum_{l=1}^n (c_{k,l}-c_{2n+1-k,l})\tilde\rho(\cos\theta_l)= -\cos\theta_k,~~k=1,\dots\, n\ ,
\end{equation}
where 
\begin{equation}
c_{kl}=\frac{\sin\theta_l}{2n+1}\left[\frac{2\epsilon_{kl}}{\cos\theta_l-\cos\theta_k}+K(Rg, \cos\theta_k,\cos\theta_l)   \right]\ {\rm and}~\epsilon_{lk}= (k-l)\mod 2\ .
\end{equation}
By choosing sufficiently large $n$ one can generate numerical solution of almost arbitrary precision. Using:
\begin{equation}
R=\left[\int_{-1}^{1}d\eta\,\tilde\rho(\eta)\right]^{-\frac12}~~{\rm and}~~\rho_1(x)=R\,\tilde\rho(x/R)\ ,
\end{equation}
one can generate an approximate numerical solution for $\rho_1(x)$. In
figure \ref{fig:1} we present the plot of the distribution $\rho_1(x)$
for range of coupling constants $0\leq g\leq 133.3$. The black dashed
curve in the figure represents the Wigner semi-circle (\ref{Wigner})
one can see the perfect fit with the data for $g=0$. The red dashed
curve fits the curve for the highest value of $g$ ($g=133.3$) and
represents the analytic expression for the distribution at large $g$
given in equation (\ref{rhoxapr}). One can see the excellent agreement
of the numerical results with the analytic analysis from the previous
subsection.
\begin{figure}[h] 
   \centering
   \includegraphics[width=14cm]{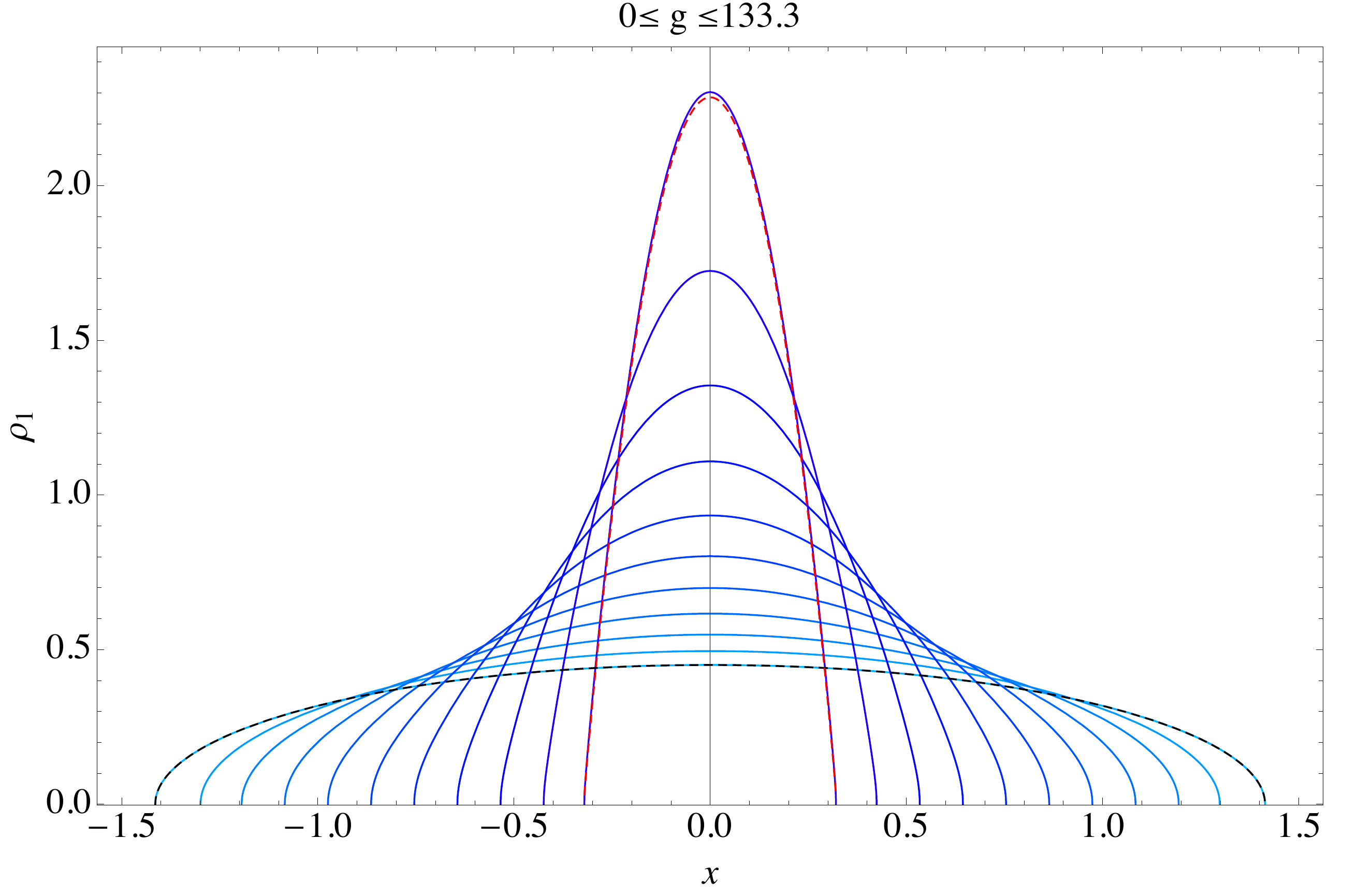}
   \caption{\small The eigenvalue density $\rho_1(x)$ for small coupling $0\leq g\leq 133.3$. The black dashed curve is the Wigner semi-circle ($g=0$) while the red dashed curve represents the approximate solution (\ref{rhoxapr}) for $g=133.3$ i.e. $Rg\simeq42.745$.}
   \label{fig:1}
\end{figure}

Let us now verify the approximate solutions for $\tilde\rho(\eta)$
obtained in the previous section. In figure \ref{fig:2} we have
presented plots of $\tilde\rho(\eta)$ for small values of the coupling
constant $0\leq Rg\leq 0.6$. The dashed red curves in figure
\ref{fig:2} represent the approximate solution for small $Rg$ from
equation (\ref{oneDpert}) while the blue ones are the numerical 
solution. One can see that the approximation is excellent
for $Rg<0.6$ and is reasonably good for $Rg=0.6$.
\begin{figure}[h] 
   \centering
   \includegraphics[width=12cm]{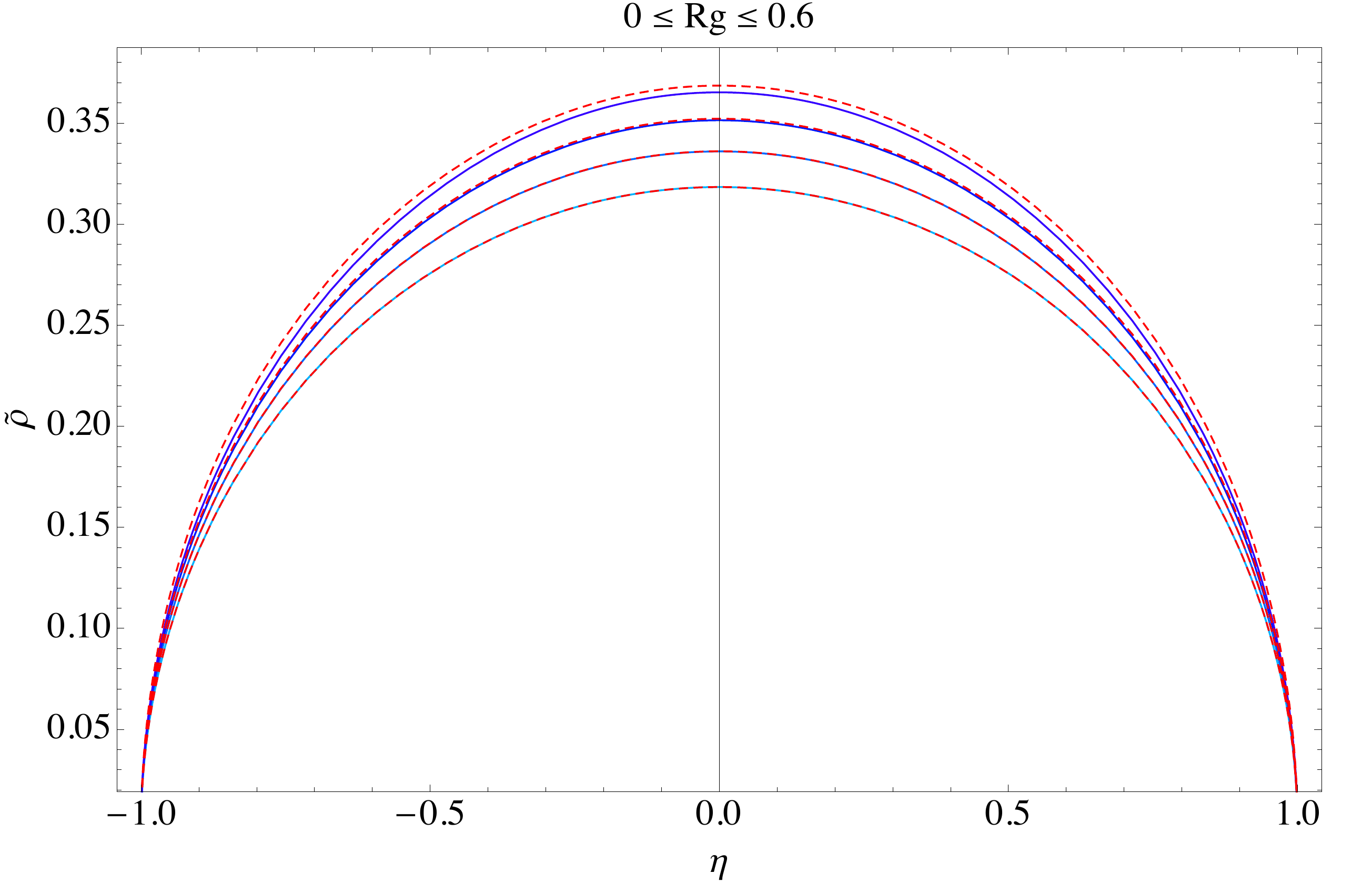}
   \caption{\small The figure shows the eigenvalue distribution $\rho_1(x)$ for 
$0\leq g\leq0.6$. The dashed red curves represent the approximate solution (\ref{oneDpert}) while the blue curve represents the numerical solution for the corresponding value of $g$. The bottom curve is the Wigner semi-circle $g=0$.}
\label{fig:2}
\end{figure}

Next we focus on the large $Rg$ regime. In figure \ref{fig:3} we
present plots of $\tilde\rho(\eta)$ for $3\leq Rg\leq 18$. The dashed
red curves represent $\tilde\rho(\eta)$ given in equation
(\ref{rhotildeLarge}). One can see how the approximation
improves as we increase $Rg$ and at $Rg=18$ it is already excellent.
\begin{figure}[h] 
   \centering
   \includegraphics[width=12cm]{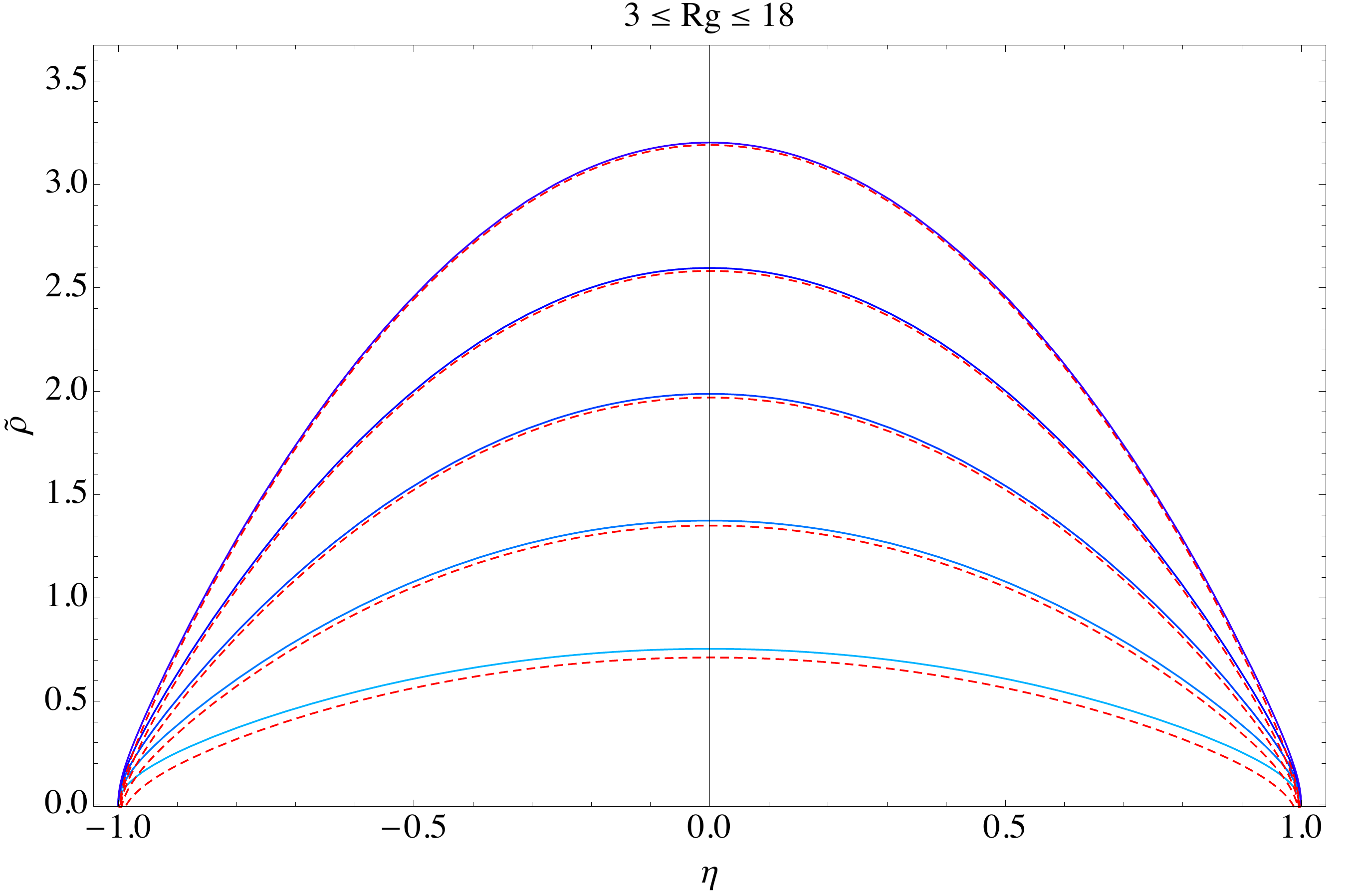}
   \caption{\small The figure shows $\tilde\rho(\eta)$ in the large $Rg$ regime $0\leq Rg\leq 18$, with the red dashed lined the theoretical expression (\ref{rhotildeLarge}) and the blue curves the numerical solution.}
\label{fig:3}
\end{figure}
\begin{figure}[h] 
   \includegraphics[width=7.5cm]{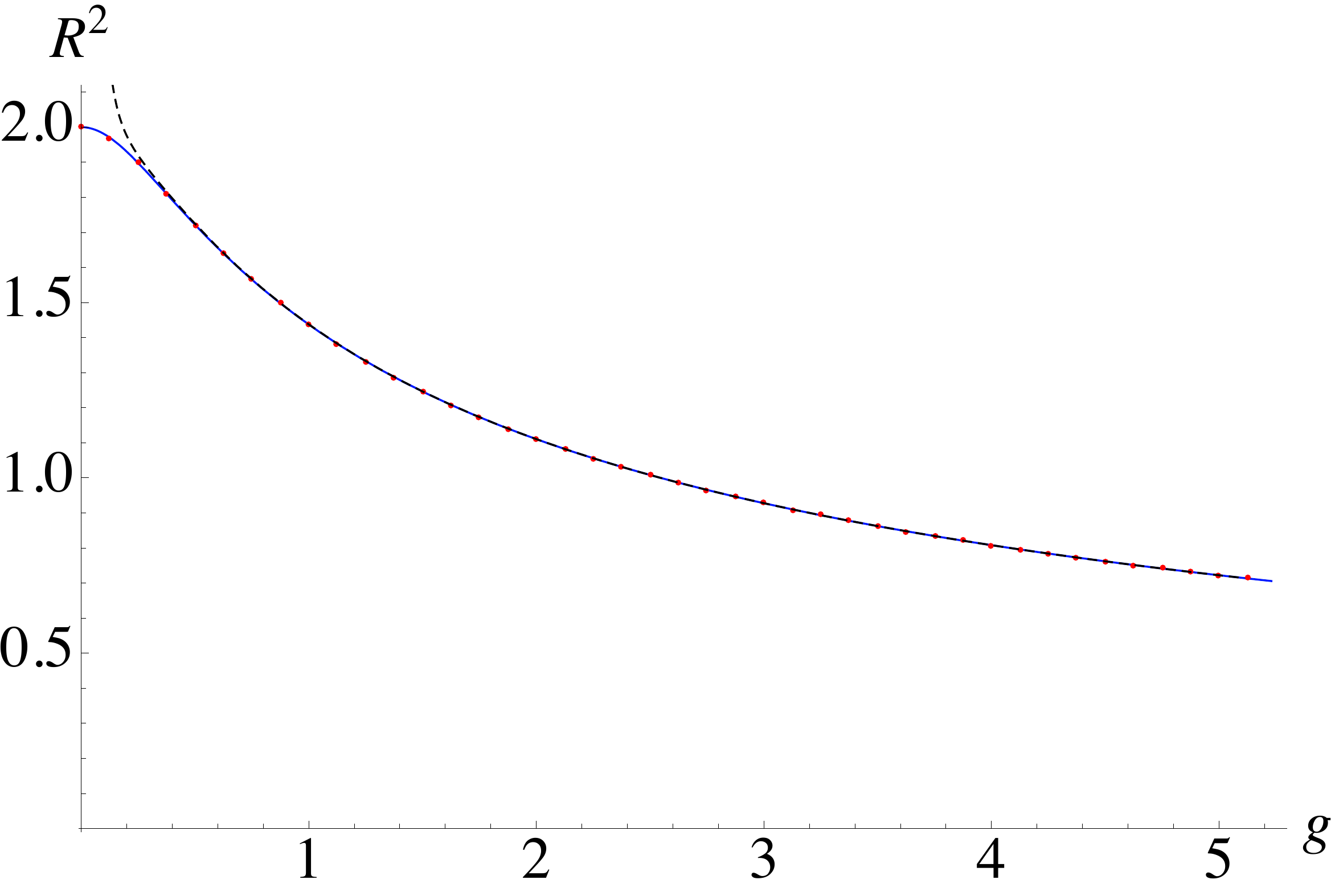}
      \includegraphics[width=7.5cm]{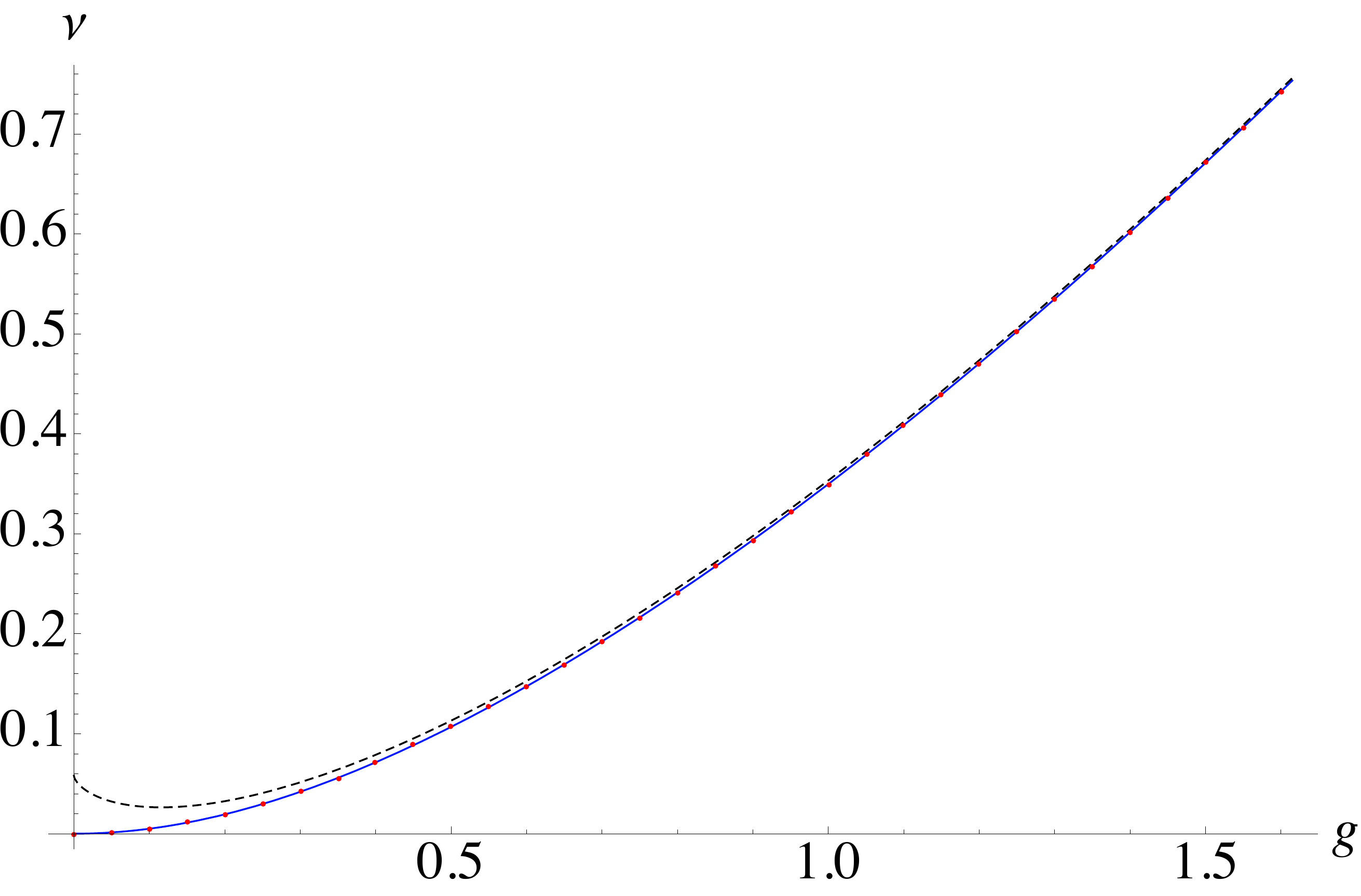}
   \caption{\small $R(g)$ and $\nu(g)$
The blue curves are the exact results (\ref{RKostov}) and (\ref{nuKostov}). 
The dashed curves are the large $g$ approximate expressions 
(\ref{RofglargegB}) and (\ref{nuapprB}), while the doted red curves 
are the numerical solution to (\ref{EQNETA}).
}
\label{fig:4}
\end{figure}

To further verify the correctness of our numerical approach we
evaluate some of the observables of the model which can be obtained in
closed form \cite{Kazakov:1998ji} (see also appendix B). In figure
\ref{fig:4} we present plots of the radius of the distribution and the
observable $\nu$ (defined in equation (\ref{nuKostov})) as a function
of the coupling constant $g$.

The blue curves in figure \ref{fig:4} represent the exact results for
the radius $R$ and the observable $\nu$ given by equations
(\ref{RKostov}) and (\ref{nuKostov}). The dashed curves are the large
$g$ approximate expressions (\ref{RofglargegB}) and (\ref{nuapprB})
one can see that for $g>1$ there is excellent agreement with the
exact result. Finally the doted red curves represent the numerical
result obtained by solving numerically the integral equation
(\ref{EQNETA}). One can observe the perfect agreement of the
numerical and the exact results.

\section{2D distribution at general coupling}

In this section we present our numerical results for the lifted 2D
distribution at general coupling. The easiest way to achieve this is
to ``lift" the distributions of the previous section.  We will describe
this in some generality.

\subsection{Lifting the distribution and Abel's integral equation}

Let us comment on the relation between rotationally invariant
distributions in different dimension, which is a generalization of the
relation (\ref{2to1}). Consider a $d-1$-dimensional distribution
$\rho_{d-1}$ obtained by reducing a $d$-dimensional rotationally
invariant distribution~$\rho_d$:
\begin{equation}
\rho_{d-1}( r )=\int\limits_{-\sqrt{R^2-r^2}}^{\sqrt{R^2-r^2}}\rho_{d}(\sqrt{r^2+z^2})dz\ .
\end{equation}
Using the change of variables: $r=\sqrt{R^2-\zeta}\, ,~z=\sqrt{\zeta-\eta}$ the  integral equation can be written as:
\begin{equation}
\rho_{d-1}(\sqrt{R^2-\zeta})=\int\limits_0^{\zeta}\frac{\rho_{d}(\sqrt{R^2-\eta})}{\sqrt{\zeta-\eta}}d\eta\ . \label{Abels}
\end{equation}
Equation (\ref{Abels}) is Abel's integral equation with a solution given by \cite{Whittaker}:
\begin{equation}
\rho_{d}(w)=\frac{1}{\pi}\frac{d}{d\eta}\int\limits_0^{\eta}\frac{\rho_{d-1}(\sqrt{R^2-t})}{\sqrt{\eta-t}}dt \Big|_{\eta=R^2-w^2}=\frac{1}{\pi\,\omega}\,\frac{d}{d\omega}\,\int\limits_R^{\omega}\,\frac{\rho_{d-1}({r})\,r}{\sqrt{r^2-\omega^2}}\,dr\ . \label{AbelSolved}
\end{equation}
It is an easy exercise to obtain the two dimensional distribution
corresponding to $\rho_1$ via equation (\ref{AbelSolved}). At zero
coupling the Wigner semicircle (\ref{Wigner}) corresponds to the a
uniform distribution $\rho=\frac{1}{2\pi}$, while at strong coupling
the parabolic distribution (\ref{parabola}) reproduces the hemisphere
distribution (\ref{hemisphere}). We will use equation
(\ref{AbelSolved}) to ``lift" the approximate solution to $\rho_1$
for general couplings.
\subsection{Numerical results}
Using equation (\ref{AbelSolved}) we can ``lift" the numerical
solution from the previous section for the one dimensional
distribution to obtain a two-dimensional rotationally invariant
distribution which in the commuting phase of the model coincides with
the joint eigenvalue distribution. Furthermore for large coupling, $g$,
we can lift the approximate expression for $\rho_1$ from equation
(\ref{rhoxapr}). Note that since $|x|\leq R\sim g^{-1/3}$ the second
term in equation (\ref{rhoxapr}) is of order $\log(g)/g^{1/3}$ and
dies out at large $g$.  The lift of the first term is part of a
semicircle of radius $(3\pi/2)^{1/3}g^{-1/3}$. The lifted distribution
at large $g$ is then given by:
\begin{equation}
\rho({x})=\begin{cases} \frac{g}{\pi^2}\left( \left(\frac{3\pi}{2g}\right)^{2/3}-x^2\right)^{1/2} +O\left(\frac{\log g}{g^{1/3}}\right)\ ,& {\rm for}~~ 0\leq x\leq R \\
0\ , &\rm{for} ~~  x > R \end{cases}\ ,\label{2Drholarge}
\end{equation}
where $R$ is given by equation (\ref{Rofglargeg}). Note that at the
boundary ($x=R$) the distribution is non-zero. Using equation
(\ref{Rofglargeg}) one can estimate the magnitude of the distribution
at the boundary:
\begin{equation}
\rho({R}) =\frac{\sqrt{\log( 96\pi^4\,g^2 )}}{2^{1/6}3^{1/3}\pi^{7/3}}{g^{1/3}} +\dots \ ,
\label{rhoBoundary}
\end{equation}
which is growing with $g$. However the magnitude of the maximum of the
distribution, at $x=0$, is $\rho(0)=(3\pi^5/2)^{1/3}\, g^{2/3}$ which
grows faster with $g$. Therefore in the limit $g\to\infty$ the shape
of the distribution approaches a hemisphere. 

In figure \ref{fig:5} we
present our numerical results for the distribution $\rho(x)$ for
different coupling constants $0\leq g\leq 12$.
\begin{figure}[h] 
   \centering
   \includegraphics[width=12.5cm]{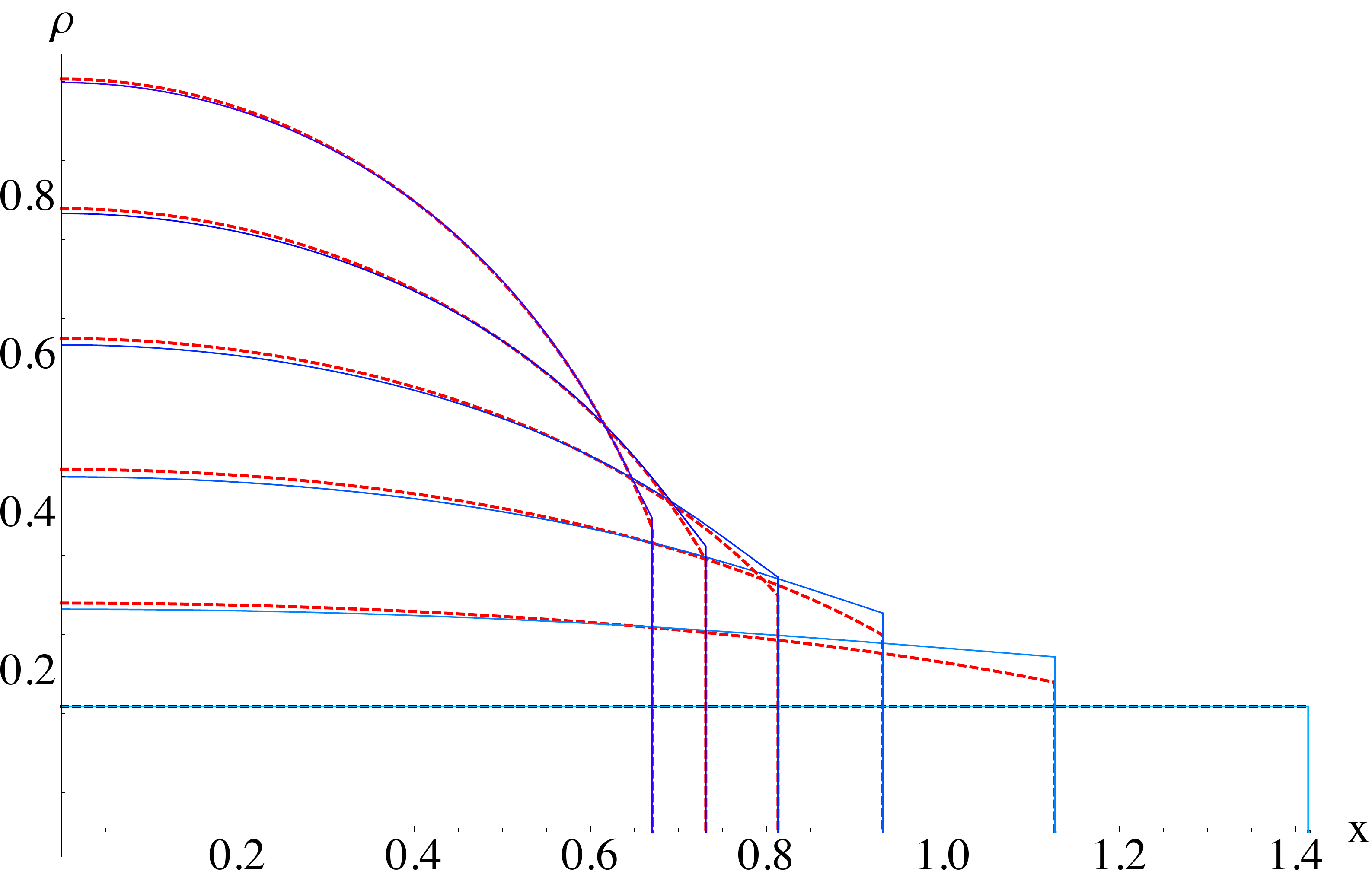}
   \caption{\small The figure shows the lifted two dimensional 
distribution as the coupling is increased. The dashed black curve (uniform distribution) is the lift of the Wigner distribution at $g=0$. The red dashed curves are the approximation (\ref{2Drholarge}) while the solid curves are the numerical results for the lifted distribution.}
   \label{fig:5}
\end{figure}
The continuous curves represent our numerical results for the lifted
distribution and the colour changes from bright blue to violet as $g$
increases. The black dashed curve represents a uniform distribution of
magnitude $1/(2\pi)$, one can see the perfect fit with the numerical
results for $g=0$ (the light blue curve). The red dashed curves
represent the approximate expression (\ref{2Drholarge}) and one can see
how the approximation improves as $g$ increases and at $g=12$ ($Rg\simeq 8$)
it is already excellent.

\section{Three matrix model realization}

Let us now consider the model (also originally introduced parenthetically 
by Hoppe \cite{Hoppe:PhDThesis1982} page 73):
\begin{equation}
{\cal Z}=\int{\cal D}X{\cal D}Y{\cal D}Ze^{-N\rm{tr}(X^2+Y^2+Z^2-i\alpha [X,Y]Z)}\ .\label{partfunct3}
\end{equation}
It is easy to verify that if one integrates out the $Z$ matrix and defines $g^2={(i\alpha)^2}/{4}$ one recovers the two matrix model (\ref{partfunct2}). 
This suggests that the model (\ref{partfunct3}) should be as solvable 
as the two matrix model. Note also that there is a global $SO(3)$ symmetry rotating the $X,Y$ and $Z$ matrices. In ref.~\cite{O'Connor:2012vr} the model (\ref{partfunct3}) was analyzed in the spirit of section 1. Namely 
the Hermitian matrices $X, Y$ and $Z$ were split into diagonal and off 
diagonal modes:
\begin{equation}
X_{ij}=x^1_i\delta_{ij}+a^1_{ij};~~Y_{ij}=x^2_i\delta_{ij}+a^2_{ij};~~Z_{ij}=x^3_i\delta_{ij}+a^3_{ij};~~\vec x_i=(x^1_i,x^2_i,x^3);~~\vec a_{ij}=(a^1_{ij},a^2_{ij},a^3_{ij})\ 
\end{equation}
and introduced an axial gauge $\vec n .\vec a=0$, where $\vec n$ is a three dimensional unit vector. After integrating out the perpendicular degrees of freedom one arrives at the following effective action:
\begin{equation}
S_{\rm{eff}}[(\vec n.\vec x)]=\frac{1}{N}\sum_{i=1}^N({\vec n.\vec x_i})^2-\frac{1}{2N^2}\sum_{i,j=1}^N\log\left[\frac{g^2 (\vec n.(\vec x_i-\vec x_j))^2}{1+g^2(\vec n.(\vec x_i-\vec x_j))^2}\right]+\frac{(N-1)}{2N}\log g^2\ .\label{eff-action3}
\end{equation}
Then considering a coarse grained approximation and varying the corresponding distribution function $\rho_3$ one arrives at the  equation:
\begin{equation}
\mu+(\vec n.\vec x)^2=\int d^3x'\rho_3(\vec x')\log\left[\frac{g^2(\vec n.(\vec x-\vec x'))^2}{1+g^2(\vec n.(\vec x-\vec x'))^2}\right]\ .\label{EQN3}
\end{equation}
Equation (\ref{EQN3}) can be averaged over a unit two-sphere (integrating both sides of the equation by $\frac{1}{4\pi}\int d\Omega_2$) to obtain:
\begin{equation}
\mu+\frac{1}{3}\vec x^2 =\int d^3 x'\,\rho_3(\vec x')\left\{\frac{-2\arctan(g|\vec x-\vec x'|)}{g|\vec x-\vec x'|}+\log\left[\frac{g^2(\vec x-\vec x')^2}{1+g^2(\vec x-\vec x')^2}\right]\right\}\label{EQN4}\ .
\end{equation}
Next we apply the Laplacian $\Delta_x$ to both side of equation (\ref{EQN4}). The result is:
\begin{equation}
1 =\int d^3x'\,\frac{\rho_3(x') }{|\vec x-\vec x'|^2(1+g^2|\vec x-\vec x'|^2)}\ , \label{EQN5}
\end{equation}
where $x=|\vec x|$ and we have used that the lifted distribution
$\rho_3$ is rotationally invariant ($\rho_3(\vec x)=\rho_3(x)$). To
obtain $\rho_3$ we need to solve the integral equation
(\ref{EQN5}). Integrating over the angular coordinates in (\ref{EQN5})
and multiplying both sides of the equation by $x$ results in:
\begin{equation}
x=\int\limits_0^R dx'\,\pi\,x'\,\rho_3(x')\,\log\left[ \frac{(x+x')^2(1+g^2(x-x')^2)}  {(x-x')^2(1+g^2(x+x')^2)}  \right]\ .\label{EQN6}
\end{equation}
If we extend the integral in equation (\ref{EQN6}) over an even
interval the integral equation can be written as:
\begin{equation}
x=\int\limits_{-R}^{R}dx'\,(-2\pi x' \rho_3(|x'|))\,K_1(g,x'-x)\ , \label{EQN7}
\end{equation}
where $K_1$ is the kernel (\ref{K1}) from section 3. Comparing
equations (\ref{EQN7}) and (\ref{eqnK1}) one arrives at the following
relation between $\rho_1$ and $\rho_3$:
\begin{equation}
\rho_3(x)=-\frac{\rho_1'(x)}{2\pi x}\ ,~~~~~~~x>0\ . \label{EQN8}
\end{equation}
In fact equation (\ref{EQN8}) can be proven in more generality. Let us
consider a rotationally invariant distribution in $d$ dimensions. The
rotationally invariant distribution in $d-2$ dimensions obtained by
integrating out two of the spacial dimensions can be obtained by
integrating over a disk:
\begin{equation}
\rho_{d-2}(x)=2\pi\,\int\limits_0^{\sqrt{R^2-x^2}}\,\rho_d(\sqrt{x^2+r^2})\,r\,dr=2\pi\,\int\limits_x^R\frac{\rho_d(\zeta)}{\zeta}\,d\zeta\ , \label{EQN9}
\end{equation}
where in the last expression we defined $\zeta=\sqrt{x^2+r^2}$. Now
after differentiating the first and the last expressions in
(\ref{EQN9}) by $x$, the integral equation for $\rho_d$ reduces to an
algebraic one which can easily be solved to obtain the analogue of
equation (\ref{EQN8}):
\begin{equation}
\rho_d(x)=-\frac{\rho_{d-2}'(x)}{2\pi x}\ ,~~~~~~~x>0\ . \label{EQN10}
\end{equation}
These considerations confirm that the procedure of averaging over $\vec
n$ in equation (\ref{EQN3}) is equivalent to lifting the one
dimensional distribution (via equation (\ref{EQN10})).

It is a straightforward exercise to ``lift" the results of section 3
to the three dimensional case using equation (\ref{EQN8}):

\subsection{3D distribution at weak coupling}
At vanishing coupling the Wigner semicircle (\ref{Wigner}) is lifted to:
\begin{equation}
\rho_3(x)=\frac{1}{2\pi^2}\,\frac{1}{\sqrt{2-x^2}}\ ,\label{EQN11}
\end{equation}
which is divergent but integrable at the boundary. In analogy with the
one dimensional case where the distribution behaves as $\sqrt{R^2-x^2}$ near
the boundary for any finite coupling, equation (\ref{EQN11}) suggests
that the lifted three dimensional distribution will 
diverge as $1/ \sqrt{R^2-x^2}$ for any finite coupling $g$.

To obtain the perturbative expression for $\rho_3$ at small $g$ it is
convenient to change variables $\eta=x/R$ and
$\tilde\rho_3(\eta)=R\,\rho_3(R\,\eta)$. The lift of equation
(\ref{oneDpert}) is then given by:
\begin{equation}
\tilde\rho_3(\eta)=\frac{1}{\sqrt{1-\eta^2}} \left[\frac{1}{2\pi^2 }+\frac{(Rg)^2}{4 \pi^2 } -\frac{\left(12 \eta ^2-5\right) (Rg)^4}{16 \pi^2 } +\frac{\left(40 \eta ^4+28 \eta ^2-31\right) (Rg)^6}{32 \pi^2 }+O\left((Rg)^8\right) \right] \ . \label{EQN12t}
\end{equation}
\subsection{3D distribution at strong coupling}

In the limit $g\to\infty$ the one dimensional distribution is
parabolic (\ref{parabola}) with radius (\ref{radius}). The lifted
three dimensional distribution is uniform (obtained in
ref.~\cite{O'Connor:2012vr}):
\begin{equation}
\rho_3(x)=\frac{g}{2\pi^2}~~\mbox{or}~~\tilde\rho_3(\eta)=\frac{Rg}{2\pi^2}\ . \label{EQN13}
\end{equation}
Note that since the model is commuting in the limit $g\to\infty$, the
lifted distribution in equation (\ref{EQN13}) is also the three
dimensional eigenvalue distribution of $X, Y, Z$.

It is straightforward to lift the correction to the distribution
$\Delta\tilde\rho$ (\ref{rhotildeLarge}) at large but finite $g$:
\begin{equation}
\tilde\rho_3(\eta)=\begin{cases}\frac{Rg}{2\pi^2}+ \frac{1}{2\pi^3(1-\eta^2)}-\frac{1}{4\pi^3\eta}\,\log\left[\frac{1-\eta}{1+\eta}\right]+O\left(\frac{\log(Rg)}{Rg}\right)\,\, & \mbox{if } |\eta|\leq1-\delta\\
0  & \mbox{if } |\eta|\geq1-\delta \end{cases} \label{EQN14}
\end{equation}
with $\delta=W(1/e)/(2\pi Rg)$ as in (\ref{rhotildeLarge}).
Finally we lift the approximate expression for $\rho_1$
(\ref{rhoxapr}) to obtain the expression for $\rho_3$:
\begin{equation}
\rho_3(x)=\frac{g}{2\pi^2}+\frac{1}{2\pi^3}\frac{\left(\frac{3\pi}{2g}\right)^{1/3}}{\left(\frac{3\pi}{2g}\right)^{2/3}-x^2}-\frac{1}{4\pi^3\,x}\log\left[\frac{\left(\frac{3\pi}{2g}\right)^{1/3}-x}{\left(\frac{3\pi}{2g}\right)^{1/3}+x} \right]+O\left(\frac{\log g}{g}\right) \label{EQN15}\ ,
\end{equation}
where $x\in(0,R)$ and $R$ is given in equation (\ref{radius}).

\subsection{3D distribution at general coupling}

In this subsection we use equation (\ref{EQN8}) to lift the interpolating solution $\tilde\rho$ from section 3.

In figure \ref{fig:5} we present a plot of the numerical solution for
the expression $\sqrt{1-\eta^2}\,\tilde\rho_3(\eta)$ for
$Rg\in[5,55]$. The color of the curves changes from blue to violet as
$Rg$ increases. The red dashed curves represent the approximate
expression (\ref{EQN14}) for
$\sqrt{1-\eta^2}\,\tilde\rho_3(\eta)$. One can see that the
approximation improves as one moves far from the boundary
$\eta=1$. One can also observe how the approximation improves as $Rg$
grows.

\begin{figure}[h] 
   \centering
   \includegraphics[width=12.5cm]{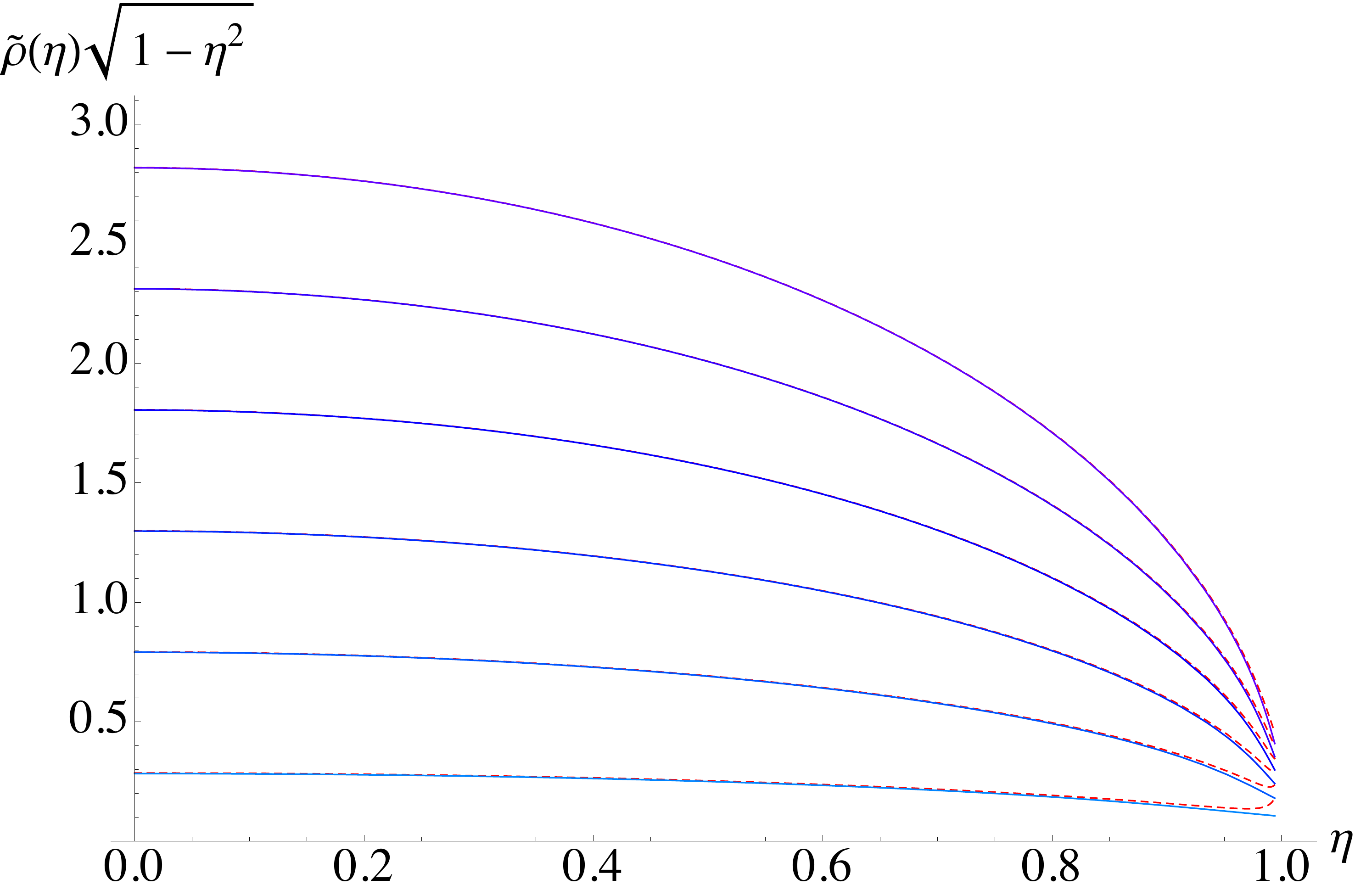}
   \caption{\small The figure shows the three dimensional rotationally invariant lifted distribution $\tilde\rho_3(\eta)$ rescaled by $\sqrt{1-\eta^2}$ 
for $5\leq Rg \leq 55$. The continuous curves are the numerical solutions to the integral equation and the dashed red curves are the approximate solution (\ref{EQN14}).}
   \label{fig:6}
\end{figure}

\section{Discussion}

We have performed a rather detailed study of Hoppe's 2-matrix model
(\ref{partfunct2}) and in particular developed both perturbative and
interpolating solutions for the eigenvalue distribution of either
matrix.  We ``lifted'' this one matrix eigenvalue distribution,
i.e. the one dimensional distribution, to rotationally invariant
distributions in both two and three dimensions. For large couplings
these lifted distributions capture the joint eigenvalue distributions
of the two and three matrix models.

We found that the two dimensional distribution does not go to zero at
the boundary but rather has a finite value, $\rho(R)$, 
given in (\ref{rhoBoundary}),  that grows as 
$\rho(R)\sim g^{\frac{1}{3}}\ln(g)$ for large $g$; see Figure \ref{fig:6}. 
This implies that the distribution lifted to a rotationally invariant 
3-dimensional distribution must diverge at the boundary for any finite $g$.
 
From Figure \ref{fig:6}, and the fact that this is the lift of the two
dimensional distribution shown in Figure \ref{fig:5}, we can deduce
that near the boundary the asymptotic behaviour of $\rho_3(x)$ is
given by
\begin{equation}
\rho_3(x)\sim\frac{g}{2\pi^2}+\frac{\rho(R)}{\pi\sqrt{(\frac{3\pi}{2g})^{\frac{2}{3}}-x^2}} .
\label{rho3asymp}
\end{equation}
The divergence\footnote{In contrast to the approximation (\ref{EQN15}),
which is not integrable at the boundary, the form (\ref{rho3asymp}) is
integrable with the divergence capturing the nature of the
distribution in the region $1-\delta\leq\frac{x}{R}\leq1$.
} of
$\rho_3(x)$ as $x$ approaches the boundary is an essential feature as
without it the limiting two dimensional distribution could not attain
a non-zero value at its boundary. This in turn means that the one dimensional 
distribution, asymptotically close to the boundary, is given by
\begin{equation}
\rho_1(x)\sim 2\rho(R)\sqrt{R^2-x^2}
\end{equation}
and the distribution crosses over to a Wigner semicircle as the
boundary is approached. 

The implication of this is that the noncommutative modes, which of
necessity are present in the model, are concentrated near the boundary
of the distribution, i.e. they are associated with the largest
eigenvalues.

The techniques used in this paper are applicable to a wide variety of
models.  The large coupling analysis of section 3, taking advantage
of the $\delta$-convergent sequence nature of the kernel, is novel and
is easily applicable to a wider class of models. Also the 
interpolation approach of section 3.3, is novel and can be adopted to more
general situations.

The models studied here have the important feature that for large couplings
the dominant configurations are commuting matrices. The fluctuations
around these commuting modes are always present and so the full
matrices never truly commute. This is in part reflected in
the divergence, as the boundary is approached, of the ``lifted'' 
three dimensional distribution. This divergence is subleading for large 
coupling but an essential feature of such models none the less.

\section{Acknowledgements} 

We thank Thomas Kaltenbrunner and Rodrigo Delgadillo-Blando for helpful discussions. The work of V. F. was supported by an INSPIRE IRCSET-Marie Curie International Mobility Fellowship.

\appendix
\section{Solving for the hemisphere distribution}
In this appendix we solve the integral equation (\ref{INTEQ}). Substituting $y(x)=x\rho(x)$ we arrive at:
\begin{equation}
f(x)=-\frac{g}{8}\left(\mu'+\frac{\vec{x}^2}{2}\right)=\int\limits_0^R\,dx' \frac{y(x')}{x+x'}K\left(\frac{2\sqrt{xx'}}{x+x'}\right)\ , \label{INTEQA}
\end{equation}
The solution to equation (\ref{INTEQA}) is given by \cite{Zabreyko,Polyanin}:
\begin{equation}
y(x)=-\frac{4}{\pi^2}\,\frac{d}{dx}\,\int\limits_{x}^{R}\,\frac{t\,F(t)\,dt}{\sqrt{t^2-x^2}}\ ,~~~F(t)=\frac{d}{dt}\,\int\limits_{0}^t\,\frac{s\,f(s)\,ds}{\sqrt{t^2-s^2}}\ . \label{SolElliptic}
\end{equation}
Substituting $f(x)$ from equation (\ref{INTEQA}) into equation (\ref{SolElliptic}) we obtain:
\begin{equation}
y(x)=x\,\frac{g}{\pi^2}\,\frac{\frac{R^2-\mu'}{2}-x^2}{\sqrt{R^2-x^2}}\ ,
\end{equation}
which implies equation (\ref{hemisphere}).
\section{Exact Results}
In this appendix we provide with slight extension some of the exact results for the model (\ref{partfunct2}) obtained in  ref.~\cite{Kazakov:1998ji}. One of the exact results of the authors of ref.~\cite{Kazakov:1998ji} was a closed form expression for the observable:
\begin{equation}
\nu=g^2\int\limits_{-R}^Rdx\,\rho_1(x)\ . \label{nuKostov}
\end{equation}
It is given by:
\begin{equation}
\nu(m)=\frac{1}{12}-\frac{K^2}{5\pi^2}\frac{10 \vartheta^2(\vartheta+m-2)+2\vartheta(6-6m+m^2)+(1-m)(m-2)}{3\vartheta^2+2(m-2){\vartheta}+1-m} \ , \label{nuKostov}
\end{equation}
where $K=K(m)$ and $\vartheta=E(m)/K(m)$ ($E $ and $K$ are the standard elliptic integrals). The elliptic modulus $m$ can be determined in terms of the coupling constant $g$ via:
\begin{equation}
g^2(m)=\frac{K^4}{3\pi^4}\left( -3\vartheta^2+2(2-m)\vartheta-(1-m)  \right)\ . \label{gKostov}
\end{equation}
Equations (\ref{nuKostov}) and (\ref{gKostov}) specify (in parametric form) the $g$ dependence of the observable $\nu$. For large $g$ one can obtain the expansion:
\begin{equation}
\nu=\frac{(12\pi)^{2/3}}{20}g^{4/3}-\frac{3}{(12\pi)^{2/3}}g^{2/3}+\left(\frac{1}{12}-\frac{1}{4\pi^2}\right)+O(g^{-2/3})\ . \label{nuapprB}
\end{equation}
Another exact result obtained in ref.~\cite{Kazakov:1998ji} relevant to our discussion is the radius of the distribution $R$ for which the authors derived the following integral presentation:
\begin{equation}
R=\frac{1}{2}\,\int\limits_{x_4}^{x_3}\,dt\,\frac{x_3-t}{\sqrt{(x_2-t)(x_1-t)(t-x_4)}}\ , \label{RKostov}
\end{equation}
where $x_1,x_2,x_3$ and $x_4$ are functions of $m$ given by:
\begin{eqnarray}
&&x_1=\frac{K^2}{g^2\pi^2}\,(2-m-2\vartheta);~~~x_2=\frac{K^2}{g^2\pi^2}\,(1-2\vartheta);\\ \nonumber
&&x_3=\frac{K^2}{g^2\pi^2}\,(3\vartheta+m-2);~~~x_4=\frac{K^2}{g^2\pi^2}\,(1-m-2\vartheta); \ .
\end{eqnarray}
 The integral (\ref{RKostov}) can be solved in closed form:
 \begin{equation}
R(m)=\frac{K(m)}{\pi\,g(m)}\,Z(\sin^{-1}\sqrt{\frac{1-\vartheta(m)}{m}}\,|\, m)\ ,\label{DenjR}
 \end{equation}
where $Z(\phi\,|\,m)$ is the standard Jacobi Zeta function.
To obtain a large $g$ expansion of $R$ we expand equation (\ref{DenjR}) near $m=1$. Using equation (\ref{gKostov}) we can obtain the expansion used in equation (\ref{Rofglargeg}):
 \begin{equation}
R=\left(\frac{3\pi}{2}\right)^{1/3}\,g^{-1/3}-\frac{2\log\,g+\log(96\pi^4)}{6\pi}g^{-1}+\frac{1}{2^{8/3}3^{1/3}\pi^{7/3}}\,g^{-5/3}+O(g^{-7/3})\ .\label{RofglargegB}
 \end{equation}

\end{document}